\documentclass[a4paper,fleqn,reqno,english,journal=cmatex,manuscript=article]{revtex4-2}
\usepackage{lmodern}
\usepackage{lmodern}
\usepackage[T1]{fontenc}
\usepackage[latin9]{inputenc}
\usepackage{color}
\usepackage{babel}
\usepackage{array}
\usepackage{float}
\usepackage{booktabs}
\usepackage{multirow}
\usepackage{amsmath}
\usepackage{amsthm}
\usepackage{amssymb}
\usepackage{graphicx}
\usepackage[unicode=true,pdfusetitle,
 bookmarks=true,bookmarksnumbered=false,bookmarksopen=false,
 breaklinks=false,pdfborder={0 0 1},backref=false,colorlinks=false]
 {hyperref}

\makeatletter

\newcommand{\lyxmathsym}[1]{\ifmmode\begingroup\def\b@ld{bold}
  \text{\ifx\math@version\b@ld\bfseries\fi#1}\endgroup\else#1\fi}

\providecommand{\tabularnewline}{\\}

\usepackage{babel}
\makeatother

\begin{document}
\title{Unlocking the Optoelectronic Potential of AGeX$_{3}$ (A = Ca, Sr,
Ba; X = S, Se): A Sustainable Alternative in Chalcogenide Perovskites}
\author{Ayan Chakravorty, Surajit Adhikari$^{*}$, and Priya Johari}
\email{ac900@snu.edu.in, sa731@snu.edu.in, priya.johari@snu.edu.in}

\affiliation{Department of Physics, School of Natural Sciences, Shiv Nadar Institution
of Eminence, Greater Noida, Gautam Buddha Nagar, Uttar Pradesh 201314,
India.}
\begin{abstract}
The quest for environmentally benign and stable optoelectronic materials
has intensified, and chalcogenide perovskites (CPs) have emerged as
promising candidates owing to their non-toxic composition, stability,
small bandgaps, large absorption coefficients. However, a detailed
theoretical study of excitonic and polaronic properties of these materials
remains underexplored due to high computational demands. Herein, we
present a comprehensive theoretical investigation of Germanium-based
CPs, AGeX$_{3}$ (A = Ca, Sr, Ba; X = S, Se), which adopt distorted
perovskite structures ($\beta$-phase) with an orthorhombic crystal
structure (space group : $Pnma$) by utilizing state-of-the-art density
functional theory (DFT), density functional perturbation theory (DFPT),
and many-body perturbation theory (GW and Bethe-Salpeter equation).
Our calculations reveal that these materials are thermodynamically
and mechanically stable, with the bandgaps calculated using G$_{0}$W$_{0}$@PBE
ranging from 0.646 to 2.001 eV $-$ suitable for optoelectronic devices.
We analyze the ionic and electronic contributions to dielectric screening
using DFPT and BSE methods, finding that the electronic component
dominates. The exciton binding energies range from 0.03 to 73.63 meV,
indicating efficient exciton dissociation under ambient conditions.
Additionally, these perovskites exhibit low to high polaronic mobilities
(1.67$-$167.65 cm$^{2}$V$^{-1}$s$^{-1}$), exceeding many lead-free
CPs and halide perovskites due to reduced carrier-phonon interactions.
The unique combination of wide tunable bandgaps, low exciton binding
energies, and enhanced charge-carrier mobility highlights AGeX$_{3}$
as a potential material for next-generation optoelectronic applications.
These compounds are stable, high-performing, and eco-friendly, showing
great promise for experimental realization and device integration. 
\end{abstract}
\maketitle

\section{Introduction:}

Inorganic-organic hybrid perovskites (IOHPs) have emerged as exceptional
contenders, achieving a remarkable surge in power conversion efficiency
(PCE) upto 26.7\% within a decade with their exceptional optoelectronic
properties such as tunable bandgaps, strong light absorption, long
carrier diffusion lengths, and remarkable defect tolerance which make
them highly attractive for next-generation devices \citep{C1-31,C1-32}.
However, large-scale commercialization remains hindered by critical
drawbacks, including lead toxicity and intrinsic instability arising
from volatile organic cations \citep{C1-33,C1-34}. These drawbacks
have spurred the search for alternative perovskite materials that
retain the superior optoelectronic properties of IOHPs while addressing
their limitations. In contrast, chalcogenide perovskites (CPs) are
gaining prominence as promising alternatives, offering superior stability,
earth-abundant constituents, and a non-toxic composition \citep{C1-35,C1-36,C1-37,C1-38}.
CPs exhibit several advantageous properties, such as optimal bandgaps,
high optical absorption coefficients, desirable charge carrier mobility,
and excellent power conversion efficiency (PCE), which collectively
enhance their suitability for optoelectronic applications \citep{C1-13,C1-27,C1-38,C1-39,C1-40}.
The successful synthesis of CPs, further underscores their viability
as next-generation semiconductors \citep{C1-36}. These attributes
firmly establish CPs as frontrunners in the pursuit of high-performance,
eco-friendly, lead-free optoelectronic materials.

Similar to halide perovskites, CPs adopt the general chemical formula
ABX$_{3}$ \citep{C1-42}, where $\lyxmathsym{\textquotedblleft}$A$\lyxmathsym{\textquotedblright}$
represents divalent alkaline-earth metals (Ca$^{2+}$, Sr$^{2+}$,
Ba$^{2+}$), $\lyxmathsym{\textquotedblleft}$B$\lyxmathsym{\textquotedblright}$
is a tetravalent metal cations (Ti$^{4+}$, Zr$^{4+}$, Hf$^{4+}$,
Sn$^{4+}$, Ge$^{4+}$), and $\lyxmathsym{\textquotedblleft}$X$\lyxmathsym{\textquotedblright}$
is a chalcogen anion (S$^{2-}$, Se$^{2-}$) . CPs primarily crystallize
in two orthorhombic phases at room temperature: the needle-like NH$_{4}$CdCl$_{3}$-type
and the distorted GdFeO$_{3}$-type, both sharing same space group
$Pnma$ (No. 62) \citep{C1-40,C1-42}. Experimentally, the distorted
phase of CaZrS$_{3}$, SrZrS$_{3}$, BaZrS$_{3}$, CaHfS$_{3}$, SrHfS$_{3}$,
and BaHfS$_{3}$ as well as the needle-like phase of SrZrS$_{3}$,
SrZrSe$_{3}$, and SrHfSe$_{3}$ have been synthesized \citep{C1-42,C1-43,C1-44,C1-45}.
Perera et al. achieved AZrS$_{3}$ (A = Ca, Sr, Ba) synthesis via
high-temperature sulfurization \citep{C1-46}, while recent studies
reported CaSnS$_{3}$ synthesis \citep{C1-47}. The phase stability
of CPs was theoretically predicted by Sun et al., and first-principles
calculations have highlighted their promising optoelectronic properties
\citep{C1-40}.

Several studies have primarily focused on Zr- and Hf-based CPs, which
exhibit exciton binding energies ranging from 0.02 to 0.26 eV \citep{C1-13,C1-27,C1-18},
comparable to or slightly higher than those of conventional HPs (0.01$-$0.10
eV) \citep{C1-48,C1-49}. However, these materials suffer from reduced
charge carrier mobility (6.84$-$77.59 cm$^{2}$V$^{-1}$s$^{-1}$)
and lower optoelectronic performance compared to conventional HPs.
These limitations arise due to prominent polaronic effects that hinder
charge transport, reducing their effectiveness for optoelectronic
applications. These challenges have steered researchers toward Ge-
and Sn-based systems, which exhibit improved charge transport and
optoelectronic properties. Recent first-principles calculations highlight
the potential of Sn and Ge as a B-site cation, with distorted SrSnX$_{3}$
(X = S, Se) perovskites emerging as promising materials based on their
optoelectronic properties \citep{C1-50,C1-51,C1-52}. Experimental
investigations have further substantiated this potential through the
successful synthesis of distorted CaSnS$_{3}$ perovskite at 500${^\circ}$C
\citep{C1-53}. Notably, needle-like CPs like BaSnS$_{3}$ and SrSnS$_{3}$
exhibit higher bandgaps (1.91$-$2.04 eV) and lower theoretical efficiency
(21.80\%) \citep{C1-27}, whereas distorted CaSnS$_{3}$ addresses
these limitations with an optimal bandgap of 1.43 eV and a significantly
higher theoretical efficiency of up to 32.45\% \citep{C1-27}, emphasizing
the superior optoelectronic performance of distorted phases.

Motivated by these findings, we extend our focus to Ge-based CPs as
the next step in CP research. As a group-IV element alongside Sn,
Ge shares similar valence and structural compatibility. Additionally,
Ge has been widely recognized as a promising material for optoelectronics
due to its proven transport properties, which have been extensively
demonstrated in high-performance electronic applications \citep{C1-54}.
Moreover, materials explored for superior optoelectronic performance
are significantly influenced by exciton formation, as efficient charge
transport relies on the thermal dissociation of excitons into free
carriers, while polaronic effects, governed by electron-phonon interactions,
dictate charge carrier dynamics and transport mechanisms. The multiple
photophysical phenomena in these materials can be understood through
electron-phonon interactions and polaron transport \citep{C1-55,C1-56}.
Despite the potential impact of these effects, a detailed theoretical
investigation of their role in Ge-based CPs remain largely unexplored,
presenting a significant research gap. To address this, we conduct
a comprehensive first-principles investigation of Ge-based CPs to
evaluate their optoelectronic properties and assess their viability
for next-generation optoelectronic applications.

Building on the promising potential of Ge-based CPs as highlighted
in the preceding discussions, we have undertaken a thorough exploration
of their electronic, optical, excitonic, and polaronic properties
of the distorted phase of AGeX$_{3}$ (A = Ca, Sr, Ba; X = S, Se)
using a combination of density functional theory (DFT) \citep{C1-57,C1-58},
density functional perturbation theory (DFPT) \citep{C1-59}, and
many-body perturbation theory techniques like GW and Bethe-Salpeter
Equation (BSE) \citep{C1-60,C1-61}. The crystal structures are optimized
with the PBE exchange-correlation functional \citep{C1-17}, and their
electronic properties are computed using the PBE xc-functional \citep{C1-17},
hybrid HSE06 functional \citep{C1-19} and G$_{0}$W$_{0}$@PBE \citep{C1-20,C1-21}
methodology. Our calculations reveal that AGeX$_{3}$ CPs exhibit
bandgaps in the optimal range for optoelectronics (0.646$-$2.001
eV), alongside exciton binding energies (0.03$-$73.63 meV) and polaronic
mobilities (upto 167.65 cm$^{2}$V$^{-1}$s$^{-1}$) that rival or
surpass conventional halide perovskites \citep{C1-48,C1-62,C1-63}.
Utilizing the DFPT technique, we compute the ionic contributions to
the dielectric function and evaluate polaronic properties using the
Fröhlich model \citep{C1-29}. Finally, some AGeX$_{3}$ compounds
exhibit very high polaron mobility, all possess low exciton binding
energies, and feature low effective masses, indicating excellent charge
carrier mobility which underscore the immense potential of Ge-based
CPs as sustainable, lead-free candidates for next-generation optoelectronic
and photovoltaic technologies.

\section{Computational Details:}

In this study, first-principles density functional theory (DFT) \citep{C1-57,C1-58}
and many-body perturbation theory (MBPT) \citep{C1-60,C1-61} calculations
were performed using the Vienna Ab initio Simulation Package (VASP)
\citep{C1-64,C1-65}. The interactions between valence electrons and
atomic cores were modeled using the projected augmented wave (PAW)
method \citep{C1-66}, with the following valence electron configurations
for the pseudopotentials: Ca (3s$^{2}$ 3p$^{6}$4s$^{2}$), Sr (4s$^{2}$
4p$^{6}$5s$^{2}$), Ba (5s$^{2}$5p$^{6}$6s$^{2}$), Ge (3d$^{10}$
4s$^{2}$ 4p$^{2}$), S (3s$^{2}$ 3p$^{4}$), and Se (4s$^{2}$ 4p$^{4}$).
The exchange-correlation (xc) interactions were treated using the
Perdew-Burke-Ernzerhof (PBE) functional within the generalized gradient
approximation (GGA) \citep{C1-17}. Structural optimization was performed
by minimizing the total energy with the conjugate gradient method
to obtain equilibrium structural parameters. A plane-wave cutoff energy
of 400 eV was used, and convergence criteria for self-consistent field
iterations and geometry relaxations were set to 10$^{-6}$ eV for
energy and 0.01 eV/$\textrm{Å}$ for Hellmann-Feynman forces on each
atom. The $\Gamma$-centered k-point grid of $8\times8\times6$ was
employed for structural optimization. PBE functionals are known to
underestimate band gaps and optical properties due to self-interaction
errors. To achieve more accurate electronic structures, hybrid Heyd--Scuseria--Ernzerhof
(HSE06) \citep{C1-19} and many-body perturbation theory (G$_{0}$W$_{0}$@PBE)
\citep{C1-20,C1-21} calculations were performed. The band structures,
including spin-orbit coupling effects, were computed with the PBE
functional. Additionally, the density of states (DOS) was calculated
using the HSE06 hybrid functional. We performed Bethe-Salpeter equation
(BSE) calculations on top of single-shot G$_{0}$W$_{0}$@PBE to accurately
determine the optical properties which explicitly accounts the electron-hole
interactions \citep{C1-23,C1-24}. The BSE calculations employed a
$\Gamma$-centered $3\times3\times2$ k-grid and 640 bands to ensure
convergence. The electron-hole kernel was constructed using 24 occupied
and 24 unoccupied states. Furthermore, ionic contributions to the
dielectric function were computed using density functional perturbation
theory (DFPT) \citep{C1-59} with an $8\times8\times6$ k-point grid.
Post-processing of elastic, electronic, and optical properties was
conducted using the VASPKIT package \citep{C1-67}.

\section{Results and Discussions:}

\subsection{\textit{Structural Properties}:}

\begin{figure}[H]
\begin{centering}
\includegraphics[width=0.4\textwidth,height=0.4\textheight,keepaspectratio]{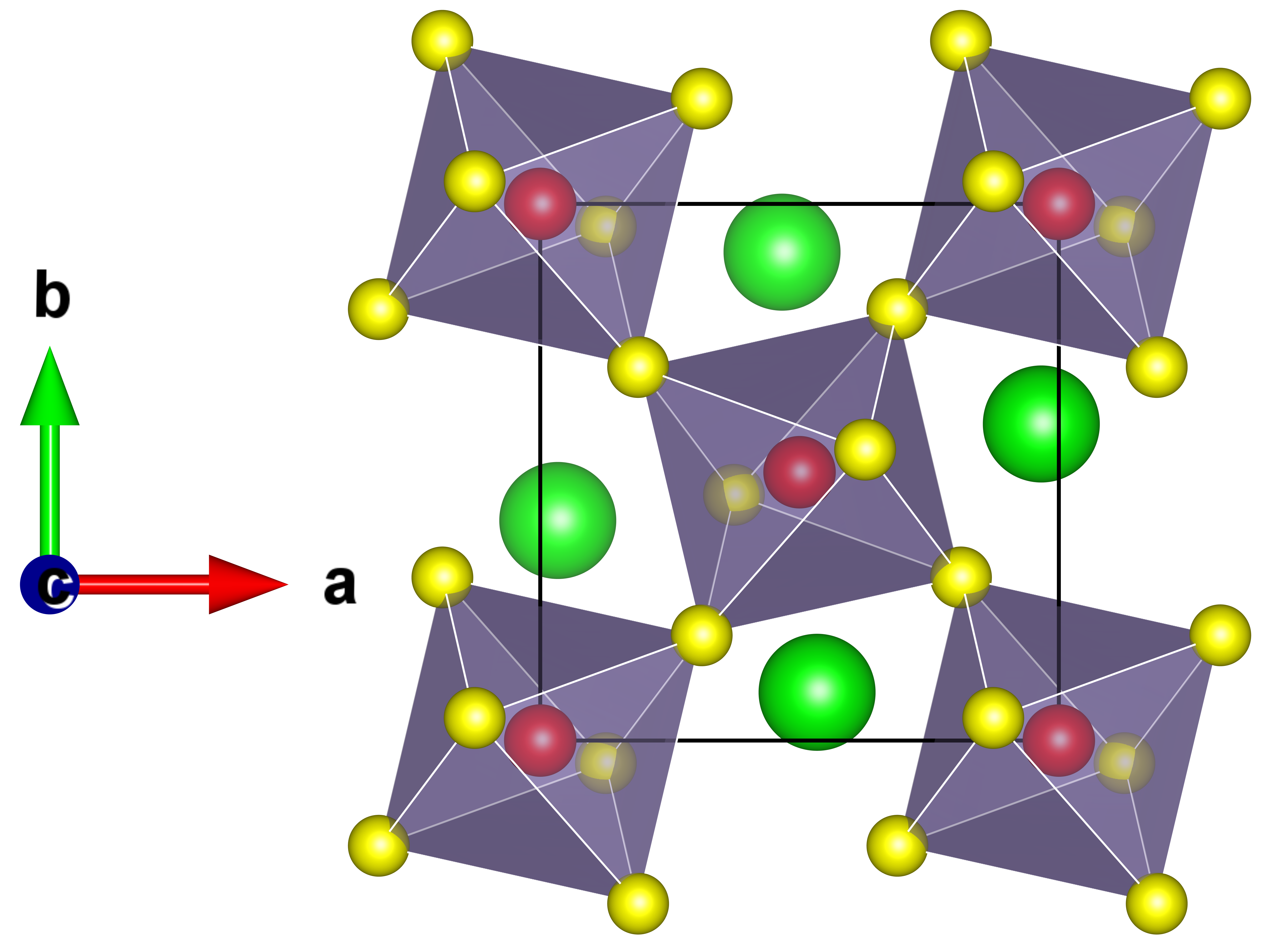} 
\par\end{centering}
\caption{\label{fig:1}Crystal structure of the orthorhombic distorted phase
for AGeX$_{3}$ (A = Ca, Sr, Ba; X = S, Se) chalcogenide perovskites.
Green, red, and yellow represent Ca/Sr/Ba, Ge, and S/Se atoms, respectively.}
\end{figure}

\subsubsection{Crystal Structure:}

In this study, we focus on the orthorhombic (space group $Pnma$)
phase of AGeX$_{3}$ (A = Ca, Sr, Ba; X = S, Se) chalcogenide perovskites
with a distorted perovskite structure. In this phase, the Ge atoms
are sixfold coordinated with the X-site anions (S or Se), forming
distorted and tilted {[}GeX$_{6}${]}$^{8-}$ corner sharing octahedras.
Each unit cell of all the compounds, comprises 20 atoms: 4 A-site
cations, 4 Ge atoms, and 12 X anions (\textcolor{black}{see}\textcolor{red}{{}
}\ref{fig:1}). The lattice parameters and bond lengths of the optimized
compounds, calculated using the PBE xc functional, are tabulated in
Table \ref{tab:1}.

The stability and structure of perovskite compounds (ABX$_{3}$) can
be qualitatively assessed using the Goldschmidt tolerance factor ($t$)
\citep{C1-1}, a widely used geometric ratio defined as: 
\begin{center}
\begin{equation}
t=\frac{r_{A}+r_{B}}{\sqrt{2}(r_{B}+r_{X})}
\end{equation}
\par\end{center}

where $r_{A}$, $r_{B}$, and $r_{X}$ are the ionic radii of the
A-site cation, B-site cation, and X-site anion, respectively. Ideal
cubic perovskites have $t$=1, corresponding to a perfect B--X--B
bond angle of 180${^\circ}$ \citep{C1-2}, while distorted perovskite
structures generally fall within 0.71 < $t$ < 0.9 \citep{C1-3}.
This structural flexibility accommodates diverse elemental combinations
at the A, B, and X-sites, enabling the exploration of unique properties
\citep{C1-4,C1-5,C1-6,C1-7,C1-8}. The ionic radii values are typically
sourced from Shannon's effective ionic radii \citep{C1-9,C1-10}.
Table S1 reveals that the $t$ values for all compounds range from
0.89$-$0.98, validating the structural stability of these distorted
chalcogenide perovskites. 
\begin{center}
\begin{table}[H]
\caption{\label{tab:1}Calculated lattice parameters, bond lengths and formation
energies of AGeX$_{3}$ (A = Ca, Sr, Ba; X = S, Se) chalcogenide perovskites.}

\centering{}{\footnotesize{}{}}%
\begin{tabular}{cccccccccccccccc}
\hline 
\multirow{2}{*}{{\footnotesize{}{}Configurations}} &  &  &  & \multicolumn{3}{c}{{\footnotesize{}{}Lattice parameter}} &  &  &  & \multicolumn{2}{c}{{\footnotesize{}{}Bond length}} &  &  &  & \multirow{2}{*}{{\footnotesize{}{}$E_{f}$ (eV/atom)}}\tabularnewline
\cline{5-7} \cline{6-7} \cline{7-7} \cline{11-12} \cline{12-12} 
 &  &  &  & {\footnotesize{}{}a ($\textrm{Å}$)}  & {\footnotesize{}{}b ($\textrm{Å}$)}  & {\footnotesize{}{}c ($\textrm{Å}$)}  &  &  &  & {\footnotesize{}{}A$-$X ($\textrm{Å}$)}  & {\footnotesize{}{}Ge$-$X ($\textrm{Å}$)} &  &  &  & \tabularnewline
\hline 
{\footnotesize{}{}CaGeS$_{3}$}  &  &  &  & {\footnotesize{}{}6.54}  & {\footnotesize{}{}6.79}  & {\footnotesize{}{}9.28}  &  &  &  & \textcolor{black}{\footnotesize{}{}2.97}{\footnotesize{} } & \textcolor{black}{\footnotesize{}{}2.46}{\footnotesize{} } &  &  &  & {\footnotesize{}{}-0.832}\tabularnewline
{\footnotesize{}{}CaGeSe$_{3}$}  &  &  &  & {\footnotesize{}{}6.99}  & {\footnotesize{}{}7.23}  & {\footnotesize{}{}9.64}  &  &  &  & \textcolor{black}{\footnotesize{}{}3.12}{\footnotesize{} } & \textcolor{black}{\footnotesize{}{}2.62}{\footnotesize{} } &  &  &  & {\footnotesize{}{}-0.736}\tabularnewline
{\footnotesize{}{}SrGeS$_{3}$}  &  &  &  & {\footnotesize{}{}6.60}  & {\footnotesize{}{}7.03}  & {\footnotesize{}{}9.49}  &  &  &  & \textcolor{black}{\footnotesize{}{}3.09}{\footnotesize{} } & \textcolor{black}{\footnotesize{}{}2.41}{\footnotesize{} } &  &  &  & {\footnotesize{}{}-0.862}\tabularnewline
{\footnotesize{}{}SrGeSe$_{3}$}  &  &  &  & {\footnotesize{}{}6.98}  & {\footnotesize{}{}7.44}  & {\footnotesize{}{}9.94}  &  &  &  & \textcolor{black}{\footnotesize{}{}3.24}{\footnotesize{} } & \textcolor{black}{\footnotesize{}{}2.56}{\footnotesize{} } &  &  &  & {\footnotesize{}{}-0.791}\tabularnewline
{\footnotesize{}{}BaGeS$_{3}$}  &  &  &  & {\footnotesize{}{}6.89}  & {\footnotesize{}{}7.12}  & {\footnotesize{}{}9.67}  &  &  &  & {\footnotesize{}{}3.33}  & \textcolor{black}{\footnotesize{}{}2.42}{\footnotesize{} } &  &  &  & {\footnotesize{}{}-0.877}\tabularnewline
{\footnotesize{}{}BaGeSe$_{3}$}  &  &  &  & {\footnotesize{}{}7.21}  & {\footnotesize{}{}7.50}  & {\footnotesize{}{}10.18}  &  &  &  & \textcolor{black}{\footnotesize{}{}3.40}{\footnotesize{} } & \textcolor{black}{\footnotesize{}{}2.58}{\footnotesize{} } &  &  &  & {\footnotesize{}{}-1.247}\tabularnewline
\hline 
\end{tabular}

\end{table}
\par\end{center}

\subsubsection{Thermodynamical Stability:}

To examine the thermodynamic stability, we calculated the formation
energy of our systems using the chemical formula: 
\begin{center}
\begin{equation}
E_{f}=\frac{E_{A_{l}Ge_{m}X_{n}}-lE_{A}-mE_{Ge}-nE_{X}}{(l+m+n)}\label{eq:2}
\end{equation}
\par\end{center}

where $E_{A_{l}Ge_{m}X_{n}}$ represents the optimized total energy
of investigated compounds and $E_{A}$, $E_{Ge}$, and $E_{X}$ are
the optimzed energies of individual A-site, Ge, and X-site (S, Se)
atoms in the crystal structure, respectively. The formation energy
for all structures are determined using Eq. \ref{eq:2} and is presented
in Table \ref{tab:1}. The results reveal that all the examined CPs
possess negative formation energies, suggesting that these materials
are thermodynamically stable under specific conditions.

\subsubsection{Mechanical stability and elastic constants:}

The elastic behavior of a material is widely recognized as a critical
factor in determining its suitability for practical device applications.
Therefore, we extended our analysis beyond crystallographic and thermodynamic
stability to include mechanical stability by calculating the second-order
elastic constants using the energy-strain approach \citep{C1-12}.
All the compounds exhibit orthorhombic crystal symmetry, requiring
nine independent elastic constants ($C_{11}$, $C_{22}$, $C_{33}$,
$C_{44}$, $C_{55}$, $C_{66}$, $C_{12}$, $C_{13}$, and $C_{23}$)
to describe their mechanical stability and elastic properties \citep{C1-12,C1-13}.
The computed $C_{ij}$ values for each CP are provided in Table S2.
All values meet the Born stability criteria \citep{C1-12} (details
in Section II of the Supplemental Material), confirming the exceptional
mechanical stability of these orthorhombic phases. Aditionally, these
elastic coefficients are used to calculate the bulk modulus ($B$),
shear modulus ($G$), Young's modulus ($Y$), and Poisson's ratio
($\nu$) for the materials \citep{C1-14,C1-15}. The brittle or ductile
nature of the materials is assessed based on the critical thresholds
for Pugh's ratio ($B/G$) and Poisson's ratio ($\nu$). Specifically,
materials with $\nu>0.26$ and $B/G>1.75$ are classified as ductile,
while those with lower values are considered brittle \citep{C1-16}.
The results indicate that the investigated CPs exhibit greater resistance
to volumetric deformation compared to shape deformation, as reflected
by the higher $B$ values relative to $G$ (See Table S2). The lower
$G$ and $Y$ values suggest a flexible nature of these materials.
The values of Pugh's ratio ($B/G$) and Poisson's ratio ($\nu$)
reveals, all the studied CPs are considered to be ductile.

\subsection{\textit{Electronic Properties:}}

Following the confirmation of structural stability, the electronic
properties of AGeX$_{3}$ (A = Ca, Sr, Ba; X = S, Se) CPs, including
band structure, total and partial density of states (TDOS and PDOS),
are analyzed to explore their potential for optoelectronic device
applications. Initially, electronic structure calculations are performed
using the GGA-PBE xc-functional \citep{C1-17}, both with and without
spin-orbit coupling (SOC). As expected for chalcogenide perovskites
\citep{C1-18}, SOC is found to have negligible impact on the bandgap,
as can be seen from Table \ref{tab:2}. However, the PBE functional
underestimates the bandgap due to self interaction error, prompted
us to use the hybrid HSE06 \citep{C1-19} functional and many-body
perturbation theory (MBPT) via the G$_{0}$W$_{0}$@PBE \citep{C1-20,C1-21}
method for more accurate bandgap estimation. The G$_{0}$W$_{0}$@PBE
calculated band structures are shown in Figure \ref{fig:2}, revealing
bandgap values ranging from 0.646 eV to 2.001 eV. For Ca- and Sr-based
compounds, the valence band maxima (VBM) are located at the $\Gamma$
(0, 0, 0) point of the Brillouin zone. In CaGeS$_{3}$ and CaGeSe$_{3}$,
the conduction band minima (CBM) are positioned between $\Gamma$
and Z (0, 0, 0.5) and between $\Gamma$ and X (0.5, 0, 0), respectively.
In SrGeS$_{3}$ and SrGeSe$_{3}$, the CBM is found at the T (0, 0.5,
0.5) point. For Ba-based compounds, BaGeS$_{3}$ exhibits a VBM at
U (0.5, 0, 0.5) and a CBM at T, while BaGeSe$_{3}$ has both the VBM
and CBM situated between T and Z, indicating an indirect bandgap.
The HSE06 (0.516$-$1.177 eV) and G$_{0}$W$_{0}$@PBE (0.646$-$2.001
eV) bandgaps fall within the range suitable for optoelectronic applications.
For example, the optimal bandgap range for solar cells is 0.9$-$1.6
eV \citep{C1-22}, making some materials particularly favorable for
photovoltaic applications, while others are better suited for alternative
optoelectronic devices. 
\begin{center}
\begin{figure}[H]
\begin{centering}
\includegraphics[width=1\textwidth,height=1\textheight,keepaspectratio]{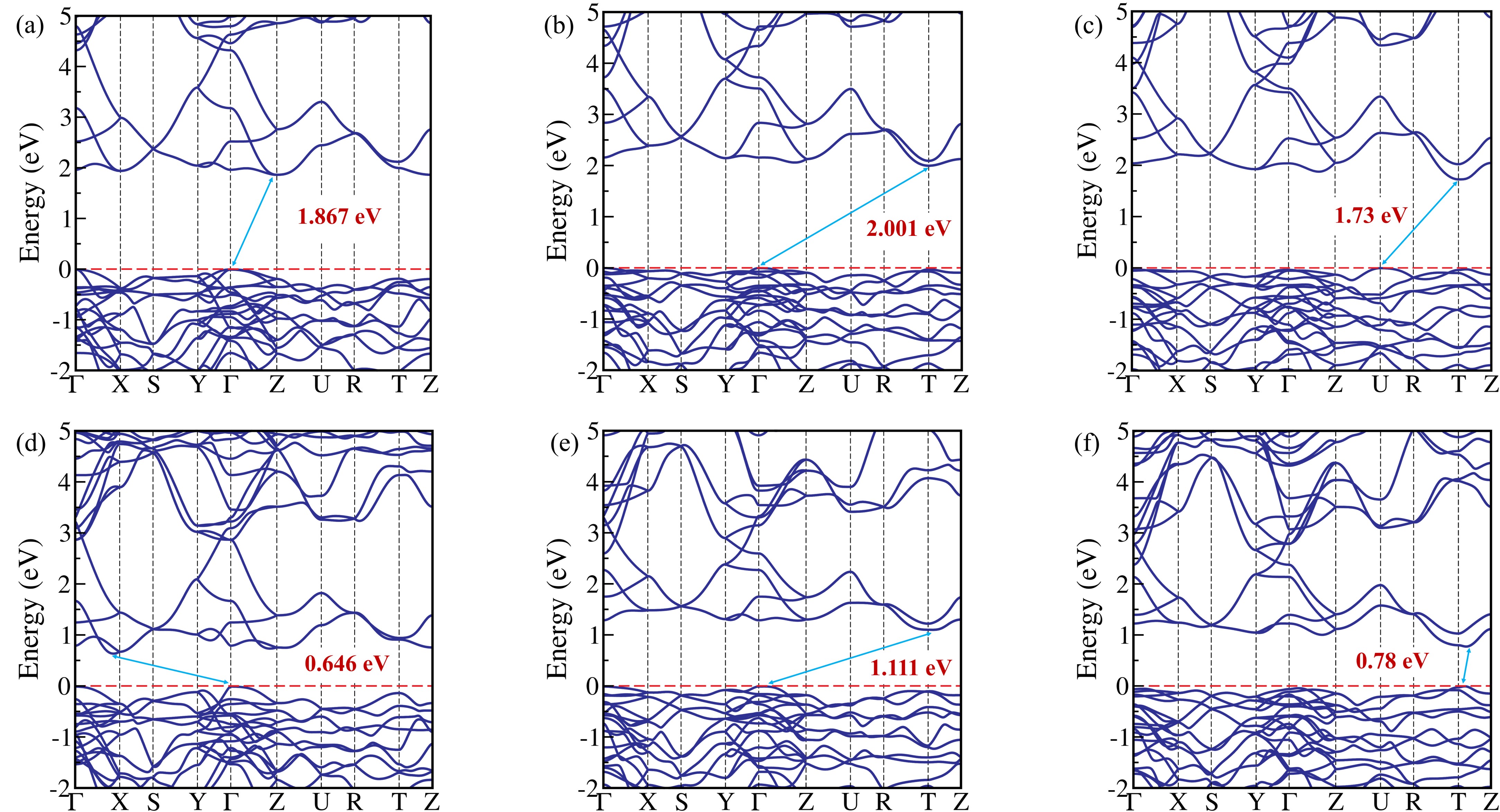} 
\par\end{centering}
\caption{\label{fig:2}Electronic band structures of AGeX$_{3}$ (A = Ca, Sr,
Ba; X = S, Se) using the G$_{0}$W$_{0}$@PBE method. The Fermi level
is set to be zero and marked by the dashed line.}
\end{figure}
\par\end{center}

The partial density of states (PDOS) and total density of states (TDOS)
for all the studied compounds, calculated using the HSE06 xc functional,
are depicted in Figure \ref{fig:3}. In the AGeS$_{3}$ (A = Ca, Sr,
Ba) CPs, VBM is predominantly derived from S-3$p$ orbitals, while
the CBM is primarily characterized by the hybridization of S-3$p$
and Ge-4$s$ orbitals. For the AGeSe$_{3}$ counterparts, the VBM
mainly originates from Se-4$p$ orbitals, with a minor contribution
from Ge-4$p$ orbitals, whereas the CBM is dominated by the hybridization
of Se-4$p$ and Ge-4$s$ orbitals. Notably, the hybridization between
Se-4$p$ and Ge-4$s$ orbitals in AGeSe$_{3}$ is more pronounced
compared to the hybridization between S-3$p$ and Ge-4$s$ orbitals
in AGeS$_{3}$. This stronger interaction in AGeSe$_{3}$ contributes
significantly to the reduction of the bandgap in selenium-based perovskites
relative to their sulfur-based counterparts, highlighting a critical
structural-electronic relationship in these materials.

To investigate the carrier transport properties of the studied perovskite
materials, we calculated the effective masses of electrons ($m_{e}^{*}$)
and holes ($m_{h}^{*}$) based on the band structure obtained from
the G$_{0}$W$_{0}$@PBE approach. The effective masses were derived
by computing the second-order derivatives of the energy dispersion
relation ($E-k$) near the band edges using the formula $m^{\ast}=\hbar^{2}\left[\partial^{2}E(k)/\partial k^{2}\right]^{-1}$,
where $\hbar$ is the reduced Planck's constant, $E(k)$ represents
the energy eigenvalue, and $k$ is the wave vector. The effective
masses are closely tied to the curvature of the conduction band minimum
(CBM) and valence band maximum (VBM)-a greater curvature in these
regions indicates lighter effective masses, which are essential for
understanding the charge carrier dynamics in these materials. From
Table \ref{tab:3}, it is evident that all compounds, except BaGeS$_{3}$,
exhibit lower electron effective masses compared to hole effective
masses, suggesting their suitability for n-type semiconductor applications.
Since these compounds are indirect band gap materials, the effective
masses were calculated at both the indirect and direct band edges.
Interestingly, BaGeS$_{3}$ shows a lower electron effective mass
at its direct band edge,\textcolor{red}{{} }\textcolor{black}{which
may result from stronger band dispersion, enhancing electron mobility
in direct band gap regions.} The lower effective masses, particularly
for electrons, indicate enhanced carrier mobility, which is beneficial
for the efficient transport of charge carriers in these materials.

\begin{table}[H]
{\scriptsize{}{}\caption{\label{tab:2}Computed bandgaps ($E_{g}$) of AGeX$_{3}$ (A = Ca,
Sr, Ba; X = S, Se) chalcogenide perovskites using the PBE/PBE-SOC,
HSE06 xc functionals, and\textbf{ }G$_{0}$W$_{0}$@PBE method.}
}{\scriptsize\par}

 \centering{}{\scriptsize{}{}}%
\begin{tabular}{cccccccc}
\hline 
\multirow{2}{*}{{\footnotesize{}{}Configurations}} & \multicolumn{3}{c}{{\footnotesize{}{}$E_{g}$ (eV) (Indirect)}} &  & \multicolumn{3}{c}{{\footnotesize{}{}$E_{g}$ (eV) (Direct)}}\tabularnewline
\cline{2-4} \cline{3-4} \cline{4-4} \cline{6-8} \cline{7-8} \cline{8-8} 
 & {\footnotesize{}{}PBE/PBE-SOC}  & {\footnotesize{}{}HSE06}  & {\footnotesize{}{}G$_{0}$W$_{0}$@PBE}  &  & {\footnotesize{}{}PBE/PBE-SOC}  & {\footnotesize{}{}HSE06}  & {\footnotesize{}{}G$_{0}$W$_{0}$@PBE}\tabularnewline
\hline 
{\footnotesize{}{}CaGeS$_{3}$}  & {\footnotesize{}{}0.548/0.519}  & {\footnotesize{}{}1.177}  & \textcolor{black}{\footnotesize{}{}1.867}{\footnotesize{} } &  & {\footnotesize{}{}0.647/0.617}  & {\footnotesize{}{}1.207}  & \textcolor{black}{\footnotesize{}{}1.965}\tabularnewline
{\footnotesize{}{}CaGeSe$_{3}$}  & {\footnotesize{}{}-/-}  & {\footnotesize{}{}0.517}  & \textcolor{black}{\footnotesize{}{}0.646}{\footnotesize{} } &  & {\footnotesize{}{}-/-}  & {\footnotesize{}{}0.567}  & \textcolor{black}{\footnotesize{}{}0.802}\tabularnewline
{\footnotesize{}{}SrGeS$_{3}$}  & {\footnotesize{}{}0.671/0.643}  & {\footnotesize{}{}1.326}  & \textcolor{black}{\footnotesize{}{}2.001}{\footnotesize{} } &  & {\footnotesize{}{}0.696/0.684}  & {\footnotesize{}{}1.345}  & \textcolor{black}{\footnotesize{}{}2.026}\tabularnewline
{\footnotesize{}{}SrGeSe$_{3}$}  & {\footnotesize{}{}0.334/0.279}  & {\footnotesize{}{}0.771}  & \textcolor{black}{\footnotesize{}{}1.111}{\footnotesize{} } &  & {\footnotesize{}{}0.427/0.366}  & {\footnotesize{}{}0.964}  & \textcolor{black}{\footnotesize{}{}1.205}\tabularnewline
{\footnotesize{}{}BaGeS$_{3}$}  & {\footnotesize{}{}0.444/0.434}  & {\footnotesize{}{}0.937}  & \textcolor{black}{\footnotesize{}{}1.730}{\footnotesize{} } &  & {\footnotesize{}{}0.473/0.448}  & {\footnotesize{}{}1.009}  & \textcolor{black}{\footnotesize{}{}1.759}\tabularnewline
{\footnotesize{}{}BaGeSe$_{3}$}  & {\footnotesize{}{}0.173/0.121}  & {\footnotesize{}{}0.516}  & \textcolor{black}{\footnotesize{}{}0.780}{\footnotesize{} } &  & {\footnotesize{}{}0.174/0.144}  & {\footnotesize{}{}0.520}  & \textcolor{black}{\footnotesize{}{}0.781}\tabularnewline
\hline 
\end{tabular}

\end{table}

\begin{center}
\begin{figure}[H]
\begin{centering}
\includegraphics[width=1\textwidth,height=1\textheight,keepaspectratio]{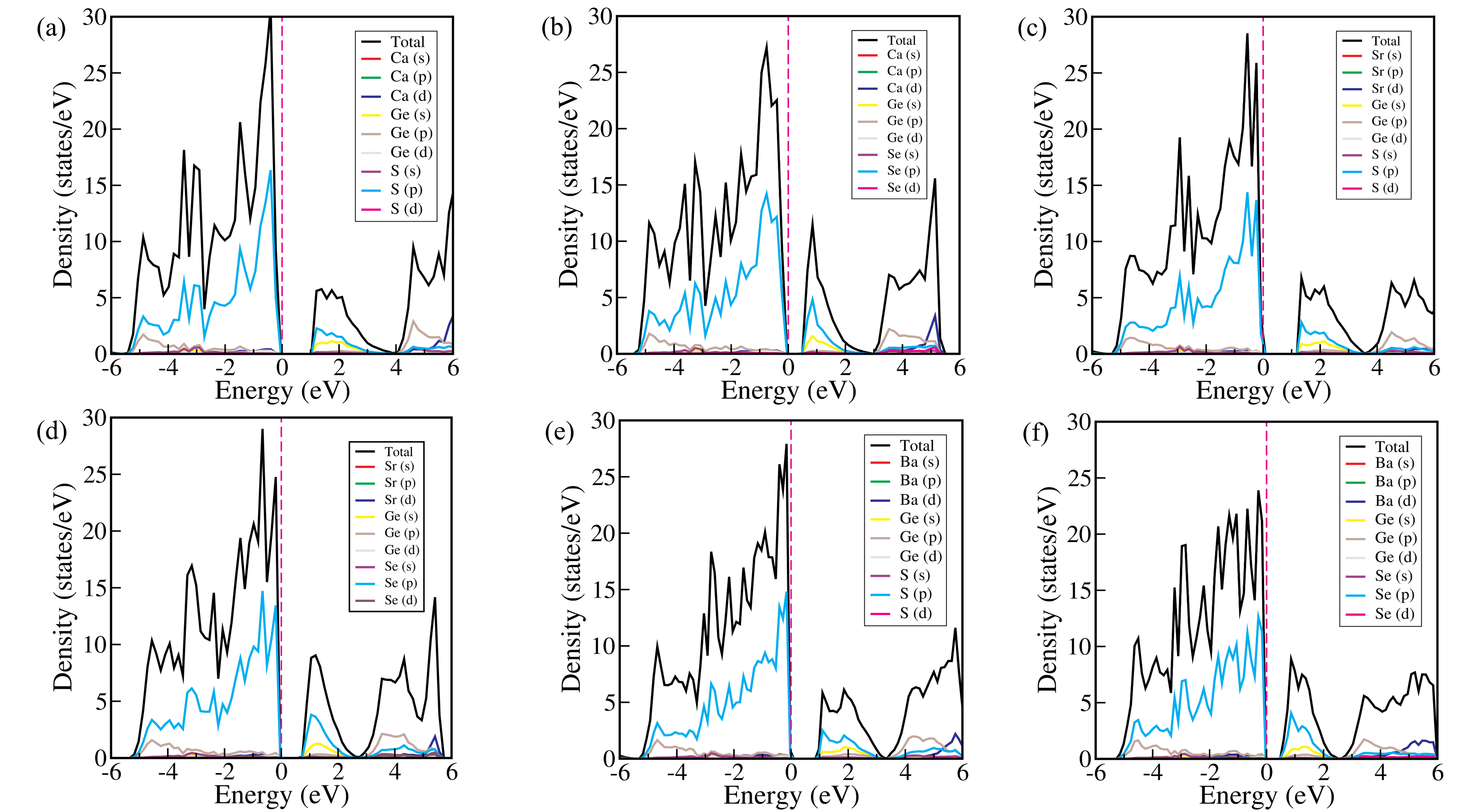} 
\par\end{centering}
\caption{\label{fig:3}Calculated electronic total density of states (TDOS)
and partial density of states (PDOS) of AGeX$_{3}$ (A = Ca, Sr, Ba;
X = S, Se) using the HSE06 xc functional. The Fermi level is set to
be zero and marked by the dashed line.}
\end{figure}
\par\end{center}

\subsection{\textit{Optical Properties:}}

A thorough analysis of a material's optical properties, such as its
dielectric function and absorption coefficient, is crucial for assessing
its suitability for optoelectronic applications. While the HSE06 functional
is effective in describing the electronic properties of the studied
compounds, it is known to provide less accurate predictions for optical
properties. To address this limitation and achieve higher accuracy,
we employed many-body perturbation theory (MBPT)-based GW-BSE calculations,
which explicitly account for electron-hole interactions. Single-shot
GW calculations were performed on top of PBE to compute the fundamental
bandgap, known to align closely with experimental techniques like
photoelectron spectroscopy (PES) and inverse photoelectron spectroscopy
(IPES) \citep{C1-20,C1-21}. Subsequently, the Bethe-Salpeter equation
(BSE) was solved on the top of G$_{0}$W$_{0}$@PBE to determine the
optical bandgap, which corresponds well with experimental optical
absorption spectroscopy \citep{C1-23,C1-24}.

\begin{table}[H]
\caption{\label{tab:3}Carrier's effective mass of AGeX$_{3}$ (A = Ca, Sr,
Ba; X = S, Se) chalcogenide perovskites\textcolor{black}{.} Here,
$m_{e}^{*}$ and $m_{h}^{*}$ represents electron and hole effective
mass (where $m_{0}$ is the rest mass of the electron), respectively.
The bold values provided in parentheses are the effective mass and
reduced mass at indirect band edges.}

\centering{}%
\begin{tabular}{cccc}
\hline 
\multirow{1}{*}{Configurations} & $m_{e}^{*}$ ($m_{0}$)  & $m_{h}^{*}$ ($m_{0}$)  & $\mu^{*}$ ($m_{0}$)\tabularnewline
\hline 
CaGeS$_{3}$  & \textcolor{black}{0.401}\textbf{\textcolor{black}{{} (0.736)}}  & \textcolor{black}{1.240}\textbf{\textcolor{black}{{} (2.230)}}  & 0.303 (\textbf{0.553})\tabularnewline
CaGeSe$_{3}$  & 0.265\textbf{\textcolor{black}{{} (0.330)}}  & 1.638\textbf{ (1.638)}  & 0.228 (\textbf{0.275})\tabularnewline
SrGeS$_{3}$  & 0.888\textbf{\textcolor{black}{{} (1.039)}}  & 1.187\textbf{ (2.841)}  & 0.508 (\textbf{0.761})\tabularnewline
SrGeSe$_{3}$  & \textcolor{black}{0.553}\textbf{\textcolor{black}{{} (0.627)}}  & \textcolor{black}{1.15}\textbf{\textcolor{black}{{} (1.284)}}  & \textcolor{black}{0.373}\textbf{\textcolor{black}{{} (0.421)}}\tabularnewline
BaGeS$_{3}$  & 0.864\textbf{ (3.777)}  & 1.118\textbf{ (2.738)}  & 0.487 (\textbf{1.587})\tabularnewline
BaGeSe$_{3}$  & 0.466\textbf{ (0.317)}  & 0.322\textbf{ (0.353)}  & 0.190 (\textbf{0.284})\tabularnewline
\hline 
\end{tabular}
\end{table}

\begin{figure}[H]
\begin{centering}
\includegraphics[width=1\textwidth,height=1\textheight,keepaspectratio]{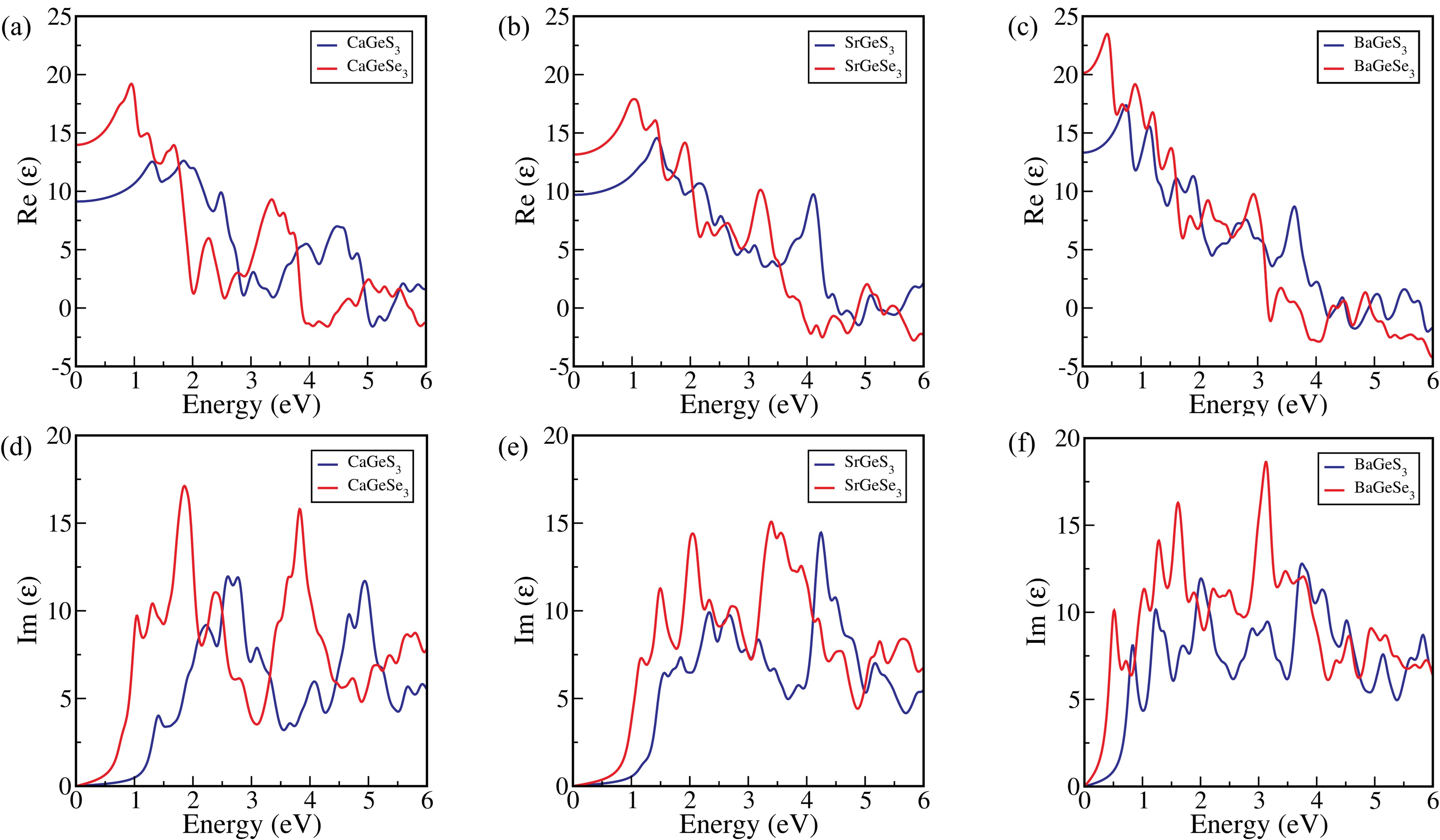} 
\par\end{centering}
\caption{\label{fig:4}(a-c) Real {[}Re ($\varepsilon$){]} and (d-f) imaginary
{[}Im ($\varepsilon$){]} parts of the dielectric function of chalcogenide
perovskites obtained using the BSE@G$_{0}$W$_{0}$@PBE method.}
\end{figure}

The optical response was further analyzed by calculating the frequency-dependent
dielectric function, $\varepsilon($$\omega$) = {[}Re($\varepsilon$){]}
+ i {[}Im($\varepsilon$){]}, where Re($\varepsilon$) and Im($\varepsilon$)
represent the real and imaginary components, respectively. Figure
\ref{fig:4} illustrates the real and imaginary components of $\varepsilon(\omega)$
evaluated using BSE@G$_{0}$W$_{0}$@ PBE. The real part of the dielectric
function {[}Re($\varepsilon$){]} reflects the material's ability
to polarize in response to an electromagnetic wave. The electronic
dielectric constant ($\varepsilon_{\infty}$), derived from this real
part, quantifies how strongly electrons in the material respond to
an electric field. A higher $\varepsilon_{\infty}$ indicates stronger
polarization, effectively screening the Coulomb interaction between
electrons and holes, thereby reducing charge carrier recombination
which is a key advantage for optoelectronic applications. Our results
reveal that the studied compounds exhibit higher $\varepsilon_{\infty}$
values compared to previously reported chalcogenide and halide perovskites
\citep{C1-13,C1-25}, highlighting their superior potential. Additionally,
$\varepsilon_{\infty}$ is observed to increase from sulfur-based
to selenium-based CPs, suggesting reduced recombination rates and
enhanced optoelectronic efficiency in the latter. The imaginary part
of the dielectric function, Im($\varepsilon$), quantifies the extent
of light energy absorbed by the material. Peaks in Im($\varepsilon$)
signify optical transitions, corresponding to the energy required
to excite electrons from the valence band to the conduction band.
The position of the first peak ($E_{o}$) denotes the optical gap,
representing the minimum photon energy necessary for electronic transitions.
The $E_{o}$ values of the studied compounds range from 0.513$-$1.420
eV, showing a redshift as the material composition changes from sulfur
to selenium. This shift toward lower energies is in line with the
reduction in quasiparticle bandgaps (see Table \ref{tab:2}) and suggests
enhanced absorption in the infrared region. Such properties make these
materials particularly suitable for applications in infrared optoelectronics
and photovoltaics.

\subsection{\textit{Excitonic Properties:}}

Excitonic properties play a pivotal role in determining the charge
separation efficiency of optoelectronic materials. Key parameters
such as exciton binding energy ($E_{B}$) and exciton lifetime ($\tau_{exc}$)
are critical for assessing their performance in practical applications.
The exciton binding energy represents the energy required to dissociate
a bound electron-hole pair into free carriers. In photovoltaic systems,
a lower $E_{B}$ enhances charge separation, contributing to improved
photoelectric conversion efficiency.

In this study, the exciton binding energy ($E_{B}$) is calculated
using the Wannier-Mott model \citep{C1-13}, which defines $E_{B}$
as:

\begin{equation}
E_{B}=\left(\frac{\mu^{\ast}}{m_{0}\varepsilon_{\mathrm{eff}}^{2}}\right)R_{\infty},
\end{equation}

where $\mu^{\ast}$ is the reduced mass of charge carriers at direct
band edges, $m_{0}$ is the electron rest mass, $\varepsilon_{\mathrm{eff}}$
represents the effective dielectric constant, and $R_{\infty}$ is
the Rydberg constant. The reduced mass ($\mu^{\ast}$) is determined
using the relation: 
\begin{center}
\begin{equation}
\frac{1}{\mu^{\ast}}=\frac{1}{m_{e}^{\ast}}+\frac{1}{m_{h}^{\ast}}
\end{equation}
\par\end{center}

To compute $E_{B}$, the effective dielectric constant ($\varepsilon_{\mathrm{eff}}$)
must be evaluated. Previous studies suggest that $E_{B}$ is influenced
by lattice relaxation effects, particularly when $E_{B}\ll\hbar\omega_{LO}$,
where $\omega_{LO}$ is the longitudinal optical phonon frequency
\citep{C1-26}. Under such conditions, $\varepsilon_{\mathrm{eff}}$
is taken as an intermediate value between the static high-frequency
dielectric constant ($\varepsilon_{\infty}$) and the low-frequency
ionic dielectric constant ($\varepsilon_{static}$). However, for
systems where $E_{B}$ significantly exceeds $\hbar\omega_{LO}$,
the electronic contribution dominates, and the ionic contribution
becomes negligible. For chalcogenide perovskites, the electronic contribution
is typically dominant \citep{C1-18,C1-13}, allowing the use of $\varepsilon_{\infty}$
as an approximation for $\varepsilon_{\mathrm{eff}}$. The calculated
$E_{B}$ values in this study, ranging from 6.38$-$73.63 meV (as
depicted in Table \ref{tab:4}), are significantly lower than those
reported for other chalcogenide perovskites \citep{C1-18,C1-27},
suggesting enhanced charge separation and, consequently, better optoelectronic
performance.

\begin{table}[H]
\caption{\label{tab:4}Calculated excitonic parameters and dielectric constants
of AGeX$_{3}$ (A = Ca, Sr, Ba; X = S, Se) chalcogenide perovskites.
Here, $\varepsilon_{\infty}$ is the electronic dielectric constant
and $E_{B}$ is the exciton binding energy (in unit meV).}

\centering{}%
\begin{tabular}{ccccc}
\hline 
Configurations  & $\varepsilon_{\infty}$  & $E_{B}$ (meV)  & $r_{exc}$ (nm)  & $|\phi_{n}(0)|^{2}$ (10$^{26}$ $m^{-3}$)\tabularnewline
\hline 
CaGeS$_{3}$  & 9.12  & 49.58  & 1.59  & 0.79\tabularnewline
CaGeSe$_{3}$  & 13.98  & 15.88  & 3.24  & 0.09\tabularnewline
SrGeS$_{3}$  & 9.69  & 73.63  & 0.88  & 4.67\tabularnewline
SrGeSe$_{3}$  & 13.16  & \textcolor{black}{29.31}  & 1.87  & 0.49\tabularnewline
BaGeS$_{3}$  & 13.32  & 37.36  & 1.45  & 1.04\tabularnewline
BaGeSe$_{3}$  & 20.14  & 6.38  & 5.61  & 0.02\tabularnewline
\hline 
\end{tabular}
\end{table}

Furthermore, using the effective dielectric constant ($\varepsilon_{\mathrm{eff}}$),
and the reduced mass of charge carriers ($\mu^{*}$), several excitonic
properties are calculated, including the excitonic radius ($r_{exc}$),
exciton lifetime ($\tau_{exc}$), and the probability density of the
electron-hole wavefunction at zero separation ($|\phi_{n}(0)|^{2}$)
is derived using the expression \citep{C1-18}: 
\begin{center}
\begin{equation}
|\phi_{n}(0)|^{2}=\frac{1}{\pi(r_{exc})^{3}n^{3}}
\end{equation}
\par\end{center}

where $r_{exc}$ represents the exciton radius, and $n$ denotes the
exciton energy level ($n$ = 1 provides the smallest exciton radius).
The excitonic radius is calculated using the following formula \citep{C1-18}: 
\begin{center}
\begin{equation}
r_{exc}=\frac{m_{0}}{\mu^{*}}\varepsilon_{\mathrm{eff}}n^{2}r_{Ry}
\end{equation}
\par\end{center}

where $m_{0}$ is the rest mass of the electron, and $r_{Ry}$ is
Bohr radius (0.0529 nm). The parameters $\varepsilon_{\mathrm{eff}}$
and $n$ are the same as defined earlier. The exciton lifetime ($\tau_{exc}$)
is inversely related to the probability of a wavefunction ($|\phi_{n}(0)|^{2}$)
for electron-hole pairs at zero charge separation (details in Section
III of the Supplemental Material). In our study, all materials exhibit
low $|\phi_{n}(0)|^{2}$ (see Table \ref{tab:4}) indicating prolonged
exciton lifetimes. Notably, Se-containing compounds show even lower
$|\phi_{n}(0)|^{2}$ due to their larger exciton radius. This extended
exciton lifetime reduces carrier recombination, enhancing conversion
efficiency and making these materials highly suitable for photovoltaic
applications.

\subsection{\textit{Polaronic Properties:}}

Electron-phonon coupling plays a essential role in shaping the physical
and chemical properties of materials, profoundly influencing carrier
dynamics and optoelectronic performance. In polar semiconductors,
such as halide and chalcogenide perovskites, the interaction of charge
carriers with longitudinal optical (LO) phonons significantly affects
carrier mobility. This interaction, driven by the macroscopic electric
field produced by LO phonons, dominates the scattering mechanism near
room temperature \citep{C1-28}. To analyze this interaction for the
studied materials, we employed the Fröhlich mesoscopic model \citep{C1-29},
which quantifies the electron-phonon coupling through the dimensionless
Fröhlich parameter ($\alpha$) with the help of the following formula: 
\begin{center}
\begin{equation}
\alpha=\left(\frac{1}{\varepsilon_{\infty}}-\frac{1}{\varepsilon_{static}}\right)\sqrt{\frac{R_{\infty}}{ch\omega_{LO}}}\sqrt{\frac{m^{*}}{m_{e}}}
\end{equation}
\par\end{center}

where $\varepsilon_{\infty}$ and $\varepsilon_{static}$ are the
electronic and static dielectric constants, respectively, while $h$
is Planck's constant, $c$ is the speed of light, and $R_{\infty}$
represents the Rydberg constant. A high $\alpha$ value ($\alpha>10$)
signifies strong coupling, while $\alpha\ll1$ indicates weak coupling
\citep{C1-29}. The calculated $\alpha$ values for AGeX$_{3}$ CPs
range from 0.78$-$5.39, indicating weak to intermediate electron
(hole)-phonon coupling. Among the compounds, BaGeSe$_{3}$ shows the
weakest coupling, while CaGeS$_{3}$ exhibits the strongest. Weak
coupling is associated with lower electron effective mass and higher
electronic dielectric constant, whereas stronger coupling correlates
with the opposite trends.

Polaron formation can lead to a reduction in the energies of electron
and hole quasiparticles (QPs). The corresponding polaron energy ($E_{p}$)
can be computed from $\alpha$ using Eq. \ref{eq:8} as, 
\begin{center}
\begin{equation}
E_{p}=(-\alpha-0.0123\alpha^{2})\hbar\omega_{LO}\label{eq:8}
\end{equation}
\par\end{center}

For CaGeS$_{3}$, CaGeSe$_{3}$, SrGeS$_{3}$, SrGeSe$_{3}$, BaGeS$_{3}$,
BaGeSe$_{3}$, the QP energy is lowered by 0.108, 0.048, 0.089, 0.042,
0.109, 0.018 eV. On comparing these values with the $E_{B}$ from
Table \ref{tab:4}, we conclude that the charge-separated polaronic
state is more stable than the bound excitons \citep{C1-18}.

The other polaron parameters, such as polaron mass ($m_{p}$) and
polaron mobility ($\mu_{p}$) are critical for evaluating the material's
optoelectronic potential. The polaron mass is calculated using the
following relation \citep{C1-11}: 
\begin{center}
\begin{equation}
m_{p}=m^{*}\left(1+\frac{\alpha}{6}+\frac{\alpha^{2}}{40}+...\right)
\end{equation}
\par\end{center}

where is $m^{*}$ is the effective mass of the charge carrier, and
$\alpha$ represents the electron-phonon coupling constant. A lower
$m_{p}$ indicates weaker carrier-lattice interactions, which is desirable
for better optoelectronic performance. Next, the polaron mobility
($\mu_{p}$) is determined with help of the Hellwarth polaron model
\citep{C1-30}: 
\begin{center}
\begin{equation}
\mu_{p}=\frac{\left(3\sqrt{\pi}e\right)}{2\pi c\omega_{LO}m^{*}\alpha}\frac{\sinh(\beta/2)}{\beta^{5/2}}\frac{w^{3}}{v^{3}}\frac{1}{K}
\end{equation}
\par\end{center}

where $\beta=hc\omega_{LO}/k_{B}T$, $e$ is the electronic charge,
$m^{\ast}$ represents the effective mass of the charge carriers,
$w$ and $v$ are associated with temperature-dependent variational
parameters (for details, see the Section V of the Supplemental Material).
$K$ is defined as follows: 
\begin{center}
\begin{equation}
K(a,b)=\int_{0}^{\infty}du\left[u^{2}+a^{2}-b\cos(vu)\right]^{-3/2}\cos(u)
\end{equation}
\par\end{center}

Here, $a^{2}$ and $b$ are evaluated as: 
\begin{center}
\begin{equation}
a^{2}=(\beta/2)^{2}+\frac{(v^{2}-w^{2})}{w^{2}v}\beta\coth(\beta v/2)
\end{equation}
\par\end{center}

\begin{center}
\begin{equation}
b=\frac{v^{2}-w^{2}}{w^{2}v}\frac{\beta}{\sinh(\beta v/2)}
\end{equation}
\par\end{center}

In this work, $\mu_{p}$ values range from 1.67 to 167.65 cm$^{2}$V$^{-1}$s$^{-1}$(see
Table \ref{tab:5}); encompassing low to very high polaron mobility.
Notably, Se-containing compounds exhibit higher mobility due to weaker
electron phonon coupling ($\alpha$), with CaGeSe$_{3}$ and BaGeSe$_{3}$
compounds showing significantly higher $\mu_{p}$ values compared
to previous reports \citep{C1-18,C1-27,C1-13}, reinforcing their
better potential for optoelectronic applications.

\begin{table}[H]
\caption{\label{tab:5}Polaron parameters corresponding to electrons ($e$)
and holes ($h$) in AGeX$_{3}$ (A = Ca, Sr, Ba; X = S, Se) chalcogenide
perovskites. Here, $\omega_{LO}$ represents the characteristic phonon
angular frequency (in unit THz), $\alpha$ represents Fröhlich interaction
parameter, $m_{p}$ represents effective mass of the polaron (in terms
of $m^{*}$), $E_{p}$ represents polaron energy (in unit meV), and
$\mu_{p}$ represents the polaron mobility (in unit cm$^{2}$V$^{-1}$s$^{-1}$),
respectively.}

\centering{}%
\begin{tabular}{cccccccccccccc}
\toprule 
\multirow{2}{*}{Configurations} & \multirow{2}{*}{$\omega_{LO}$ (THz)} &  & \multicolumn{2}{c}{$\alpha$} &  & \multicolumn{2}{c}{$m_{p}/m^{*}$} &  & \multicolumn{2}{c}{$E_{p}$ (meV)} &  & \multicolumn{2}{c}{$\mu_{p}$ (cm$^{2}$V$^{-1}$s$^{-1}$)}\tabularnewline
\cmidrule{4-5} \cmidrule{5-5} \cmidrule{7-8} \cmidrule{8-8} \cmidrule{10-11} \cmidrule{11-11} \cmidrule{13-14} \cmidrule{14-14} 
 &  &  & $e$  & $h$  &  & $e$  & $h$  &  & $e$  & $h$  &  & \multirow{1}{*}{$e$} & $h$\tabularnewline
\midrule 
CaGeS$_{3}$  & 3.01  &  & 3.10  & 5.39  &  & 1.76  & 2.62  &  & 38.62  & 68.97  &  & 13.13  & 1.68\tabularnewline
CaGeSe$_{3}$  & 3.29  &  & 1.05  & 2.33  &  & 1.20  & 1.52  &  & 14.89  & 33.55  &  & 102.66  & 8.43\tabularnewline
SrGeS$_{3}$  & 4.21  &  & 1.91  & 3.16  &  & 1.41  & 1.78  &  & 33.23  & 55.81  &  & 15.39  & 3.16\tabularnewline
SrGeSe$_{3}$  & 3.01  &  & 1.41  & 2.03  &  & 1.28  & 1.44  &  & 17.21  & 24.97  &  & 43.39  & 13.54\tabularnewline
BaGeS$_{3}$  & 3.67  &  & 3.76  & 3.23  &  & 1.98  & 1.80  &  & 59.01  & 50.37  &  & 1.67  & 2.90\tabularnewline
BaGeSe$_{3}$  & 2.56  &  & 0.78  & 0.82  &  & 1.15  & 1.15  &  & 8.66  & 9.11  &  & 167.65  & 142.45\tabularnewline
\bottomrule
\end{tabular}
\end{table}

\section{Conclusions:}

To summarize, this study comprehensively explores the ground- and
excited-state properties of Ge-based chalcogenide perovskites AGeX$_{3}$
(A = Ca, Sr, Ba; X = S, Se), utilizing advanced DFT, DFPT, and MBPT
(GW, BSE) methods. The materials demonstrate structural, thermodynamical,
and mechanical stability, with band gaps calculated using the G$_{0}$W$_{0}$@PBE
method, ranging from 0.646$-$2.001 eV. All the compounds having high
charge carrier mobility, owing to small effective masses. The exciton
binding energies (0.03$-$73.63 meV) are comparable to or lower than
other chalcogenide perovskites, with a dominant electronic contribution
to dielectric screening. Polaron mobility for electrons (1.67$-$167.65
cm$^{2}$V$^{-1}$s$^{-1}$) and holes (1.68$-$142.45 cm$^{2}$V$^{-1}$s$^{-1}$)
further supports their excellent optoelectronic potential. These materials
exhibit promising optical properties, making them ideal for a wide
range of optoelectronic applications, including light-emitting diodes,
photodetectors, and solar cells. Overall, the findings reveal the
remarkable promise of Ge-based chalcogenide perovskites in advanced
optoelectronic devices, offering a robust pathway toward efficient
and sustainable next-generation technologies. 
\begin{acknowledgments}
A.C. would like to acknowledge the Shiv Nadar Institution of Eminence
(SNIoE) for funding and support. S.A. would like to acknowledge the
Council of Scientific and Industrial Research (CSIR), Government of
India {[}Grant No. 09/1128(11453)/2021-EMR-I{]} for Senior Research
Fellowship. The authors acknowledge the High Performance Computing
Cluster (HPCC) `Magus' at SNIoE for providing computational resources
that have contributed to the research results reported within this
paper. 
\end{acknowledgments}

 \bibliographystyle{apsrev4-2}
\bibliography{refs}

\end{document}